\documentclass{aastex}
\usepackage{emulateapj5,graphicx}

\shorttitle{Ionizing Efficiency of the First Stars}
\shortauthors{Venkatesan \& Truran}

\begin{document}
\submitted{Accepted for publication in ApJLetters, v. 594 (Sept. 1, 2003)}

\title{The Ionizing Efficiency of the First Stars}

\author{Aparna Venkatesan$^{1,2}$}
\author{James W. Truran$^3$}
\altaffiltext{1}{CASA, Department of Astrophysical and Planetary Sciences,
 University of Colorado, UCB 389, Boulder, CO 80309-0389}
\altaffiltext{2}{NSF Astronomy and Astrophysics Postdoctoral Fellow}
\altaffiltext{3}{Department of Astronomy and Astrophysics, 5640 S. Ellis
  Ave., University of Chicago, Chicago, IL 60637}
\vspace{0.1in}


\begin{abstract}

We investigate whether a single population of first stars could have
influenced both the metal enrichment and reionization of the high-redshift
intergalactic medium (IGM), by calculating the generated ionizing radiation
per unit metal yield as a function of the metallicity of stellar
populations. We examine the relation between the ionizing radiation and
carbon created by the first stars, since the evidence for the widespread
enrichment of the IGM at redshifts $z \sim 3-4$ comes from the detection of
C~IV absorption.  We find that the number of ionizing photons per baryon
generated in association with the detected IGM metallicity may well exceed
that required for a late hydrogen reionization at $z \sim 6$, by up to a
factor of 10--20 for metal-free stars in a present-day initial mass
function (IMF).  This would be in agreement with similar indications from
recent observations of the microwave background and the high-$z$ IGM. In
addition, the contribution from intermediate-mass stars to the total metal
yield, neglected in past works, substantially impacts such
calculations. Lastly, a top-heavy IMF is not necessarily preferred as a
more efficient high-$z$ source of ionizing radiation, based on
nucleosynthetic arguments in association with a given level of IGM
enrichment.

\end{abstract}


\section{Introduction}

The first stars are expected to be cosmologically significant from at least
two sets of observations of the high-redshift ($z$) intergalactic medium
(IGM): (1) hydrogen reionization at $z \ga 6$ \citep{becker}, and (2) the
persistent metallicity ($Z \sim 0.003 Z_\odot$) seen in Ly$\alpha$ clouds
up to $z \sim 5$ \citep{songcow, song01}.  In principle, there should be a
strong correlation between the integrated stellar output of metals and
ionizing radiation (\citealt{gs96}, and references therein).  Examining the
connection between these two effects is critical for several problems in
cosmology, including the degree to which a {\it single population of early
stars} could have influenced both reionization and IGM metal pollution, and
the accurate calculation of the IGM metallicity, $Z_{\rm IGM}$, from
observations of metal line absorption, which depends on the assumptions
made for the ionizing photon background (e.g., \citealt{girshull97}). These
issues are also pertinent to other fields that are constrained by the
stellar history of the baryons, such as theories for baryonic dark matter
candidates \citep{fields}.

Current spectroscopic data imply that H~I reionization may occur not far
beyond $z \sim 6$ \citep{becker}, and that He~II reionization occurs at $z
\sim 3$ (\citealt{kriss}, and references therein). However, recent data
suggest a higher level of ionizing radiation at $z \ga$ 6--9 than would
occur in a scenario with ``late'' H~I reionization at $z \sim$ 6. These
include observations of: the microwave background from the $WMAP$
satellite, indicating a high Thomson optical depth of $\sim 0.17 \pm 0.04$
\citep{kogut} or early reionization at $z \sim 17 \pm 4$; the evolution of
Ly-limit systems and the IGM ionizing background \citep{escude03}; and the
observed redshift evolution of the Fe/Mg ratios in QSO emission-line
regions \citep{hamferland}. Although these data are by no means conclusive,
they indicate that the level of stellar activity at $z \ga 9$ is
potentially significant, and that the reionization history of the IGM could
be relatively complex, with periods of extended or double reionization for
H~I and/or He~II \citep{vts,cen03,wyithe,huihaiman}. Given the current
best-fit cosmological parameters and the associated small-scale power
available to the reionizing sources \citep{spergel}, an unusually high
conversion efficiency of baryons to ionizing radiation may be required
\citep{somerville1,ciardi03}.

At present, it is believed that almost all of the baryons at $z$ $\ga 3$
reside in the IGM, with an average enrichment of $Z_{\rm IGM} \sim$
$10^{-2.5} Z_\odot$ at $z \sim 3$ \citep{ellison}. This is detected
primarily through C~IV absorption down to the lowest column density
systems, corresponding to the ``true IGM''; there is also some evidence for
widespread O~VI enrichment in this IGM component \citep{schaye}.  From
this, one can roughly estimate the fraction of baryons that went into the
first stars that made the IGM metals, and the associated number of ionizing
photons per baryon \citep{mirees97,hl97}.  However, there are two potential
problems that we face in trying to constrain the effects of high-mass stars
consistently through the degree of IGM carbon enrichment. First, the
photons relevant for reionization come from the massive stars in the
stellar initial mass function (IMF; stellar masses $\ga$ 10 M$_\odot$),
while carbon is produced dominantly by longer-lived intermediate-mass stars
(IMSs; $\sim$ 2--6 M$_\odot$). These stellar products may not be mutually
constraining if the IMF was different in the past. Second, for burst-driven
star formation (not continuous), most of the carbon is produced $\sim$ 0.5
Gyr after the starburst, an order of magnitude in time {\it after} the
massive stars' Type II SNe, which occur on timescales of a few tens of Myr
and which could drive galactic winds. It then becomes unclear how the
carbon is expelled from individual halos and distributed ubiquitously in
the IGM by $z \sim$ 3-5 \citep{song01}.

In this Letter, we present calculations of the stellar ionizing efficiency
as a function of stellar metallicity, $Z_\star$, and of the generated
metals. We specifically examine the relation between the integrated
ionizing radiation and the output in carbon and oxygen as: (1) these metals
appear to be widely detected in the IGM, but are primarily the products
(and hence probes) of different parts of the IMF, and (2) carbon and oxygen
are easily observed at typical IGM temperatures at $z \sim $ 2--4. The
point that carbon is primarily the product of IMSs while massive stars
dominate the ionizing radiation is not new to this work, and has been
demonstrated extensively in the past (see respectively, e.g.,
\citealt{ibentruran}, and \citealt{tsv} and references therein). Our main
goals here are to investigate the implications of these stellar trends for
cosmology, particularly for the reionization and metal enrichment of the
high-$z$ IGM. We focus on stellar nucleosynthetic arguments, from which we
can derive the minimum number of ionizing photons that must have been
generated in association with the observed levels of IGM enrichment at $z
\sim 2-5$, including cases with the significant metal yield from IMSs.

\section{Relating Stellar Radiation and Metal Yields}

The first generations of stars are expected to be metal-free in composition
(Pop III), and rely more heavily on the p--p chain initially than on the
more efficient CNO cycle for their thermonuclear fuel source. Consequently,
they are hotter and emit significantly harder ionizing radiation relative
to their finite-$Z_\star$ counterparts \citep{bromm, tsv, sch02, sch03}.
These objects could play an important role in either the single or extended
reionization of H~I and He~II.  The gain in ionizing radiation with
metal-free stars depends on the assumed IMF. Recent theoretical studies
indicate that the primordial stellar IMF may have been top-heavy
\citep{abel, bromm02}, leading predominantly to stellar masses $\ga$ 100
$M_\odot$.

As noted earlier, for a Salpeter IMF, all of the ionizing radiation and
most of the metals, particularly the alpha-elements (e.g., oxygen),
originate from the short-lived massive stars.  With decreasing $Z_\star$ of
the stellar population, the massive stars generate greater ionizing
radiation and lower post-SN metal yield \citep{heger02, ww95}. However, the
opposite trend holds for the IMSs: a star of lower $Z_\star$ has greater
metal yield \citep{vangroen}.  Such constraints on stellar activity
therefore alter with the average $Z_\star$ of starforming regions and hence
with redshift.

As a diagnostic of the relation between the generated ionizing photons
and metals from stars, \citet{madshull} introduced the conversion
efficiency, $\eta_{\rm Lyc}$, of energy produced in rest mass of metals,
$M_Z c^2$, to the energy released in the H-ionizing continuum, defined as
follows:
\begin{equation}
\eta_{\rm Lyc} \equiv E (h\nu \geq 13.6 \, {\rm eV})/(M_Z c^2) 
\end{equation}

\citet{madshull} found $\eta_{\rm Lyc}$ to have a relatively constant
value of 0.002\footnote{More recent unpublished calculations indicate
that $\eta_{\rm Lyc} \sim 0.003$ for $Z_\star = Z_\odot$ (J.~M. Shull
2003, private communication).} for solar values of $Z_\star$, and to
be roughly independent of the stellar IMF's slope. This quantity
can be used to calculate the number of ionizing photons per baryon in
the universe generated in association with the observed IGM
metallicity, $N_{\rm Lyc}/N_{\rm b}$, defined as the ratio of the
energy of the Lyc photons per baryon to the average energy of a Lyc
photon \citep{mirees97}. For solar-$Z_\star$ stars where $\eta_{\rm
Lyc} = 0.002$, we have:
\begin{eqnarray}
N_{\rm Lyc}/N_{\rm b} & = &  (E_{\rm Lyc}/N_{\rm b})/\langle E_{\rm Lyc}
\rangle \nonumber \\ & = &
(\eta_{\rm Lyc} \langle Z_{\rm IGM} \rangle m_p c^2)/\langle E_{\rm Lyc} \rangle \nonumber \\ & \sim &
0.002 \times 10^{-4} \times \, (1 \, {\rm GeV}/20 \, {\rm eV}) \sim 10 .
\end{eqnarray}

This implies that, on average, 10 stellar ionizing photons per baryon were
generated in association with the synthesis of the observed $Z_{\rm IGM}
\sim 10^{-4}$, a reasonable value for a number of reasons.  Given the
observed decline in the space density of bright QSOs at $z \ga$ 3,
high-redshift star formation is thought to have played a significant, if
not dominant, role in reionization. Furthermore, although $N_{\rm
Lyc}/N_{\rm b}$ need only equal about unity for reionization, the effects
of high-$z$ recombinations in the IGM and individual halos likely boost the
required $N_{\rm Lyc}/N_{\rm b}$ to values close to about 10 (see
\citealt{somerville1}).

However, $\eta_{\rm Lyc}$ is strongly dependent on $Z_\star$ and the IMF
mass range, an effect that has not been studied until recently
\citep{sch02}.  Thus, the above calculation of $\eta_{\rm Lyc}$ does not
always lead to $N_{\rm Lyc}/N_{\rm b}$ $\sim$ 10. Pop III stars, having no
metals, are hotter and emit harder radiation. When combined with the trend
of lower metal yields with decreasing $Z_\star$, we would expect a net
decline in $\eta_{\rm Lyc}$ with rising $Z_\star$.  This has been recently
demonstrated for a few specific cases by \citet{sch02}. However, the
significant contribution by IMSs to the overall metal yield, particularly
through carbon, has been neglected by past works \citep{sch02, madshull},
which assumed that massive stars were the dominant sources of the
metals. Including the IMS contribution has a non-negligible impact on the
values of $\eta_{\rm Lyc}$ as well as on $N_{\rm Lyc}/N_{\rm b}$, as we
show below.

A concern in relating the ionizing photon and nucleosynthetic output of
stellar populations is finding an overlap amongst the input $Z_\star$ and
assumptions in the literature's calculations. At present, the ionizing
photon spectrum for metal-free massive stars appears to be consistent
amongst several works to within $\sim$ 10\%
\citep{sch03,tsv,bromm}. However, the stellar yields are not entirely
convergent yet, with variations of up to factors of a few (see, e.g,
\citealt{limchieffi}).  We consider three values of $Z_\star$: 0, 0.001 and
0.02 (i.e., solar), and two separate mass ranges, 1--100 $M_\odot$ and
100--1000 $M_\odot$. From here onwards, we refer to the former mass range
as a present-day IMF and to the latter as very massive stars (VMSs) or,
equivalently, a top-heavy IMF. Although such clean distinctions may not
exist for the first stars, we do this in order to differentiate the
signatures of stars in these two IMFs. The metal yields are taken from
\citet{vangroen} for stellar masses $<$ 8 M$_\odot$ and from \citet{ww95}
for the 8--40 M$_\odot$ mass range. Note that for $Z_\star=0.001$ massive
stars, we use the $Z=0.1 Z_\odot$ case from \citet{ww95}. Since there are
no detailed yields to date for $Z_\star=0$ IMSs, we use those for
$Z_\star=0.001$ IMSs, which are not substantially different (A. Chieffi
2002, private communication). For $Z_\star=0$ VMSs, only the stars of mass
140--260 $M_\odot$ avoid complete collapse into a black hole
\citep{heger02} and contribute to the nucleosynthetic output.  We note that
the term yield in this paper connotes the total ejected mass in individual
elements or in metals, and not the net yield after accounting for the
original metal composition of the star.

We do not consider main-sequence mass loss in any of our cases below; this
is unlikely to be important for low-$Z_\star$ stellar populations (see
\citealt{tsv} on this point). We assume the stellar IMF to have the form,
$\phi(M) = \phi_0 M^{-\alpha}$, and the IMF slope, $\alpha$, to be the
Salpeter value of 2.35, since small slope variations have been shown to
make little difference \citep{gs96,madshull,sch02,sch03}. All the cases
here are normalized over their respective mass ranges as, $ \int dM M
\phi(M) = 1 $.

For the ionizing spectra, we use \citet{bromm} and \citet{sch03} for a
top-heavy IMF, \citet{tsv} for 1--100 M$_\odot$, $Z_\star=0$ stars, and
\citet{leitherer} for non-zero $Z_\star$ cases. We note that the {\it
time-integrated} ionizing photon number from VMSs happens to be roughly the
same as that from $Z_\star=0$ stars in an IMF over 1--100 $M_\odot$,
despite the greatly boosted ionizing {\it rate} from VMSs, owing to their
brief lifetimes.  The stellar emission rate of ionizing photons drops to
less than 1\% of its initial value after 30 Myr for a present-day IMF
\citep{vts}, but takes only about 3.5 Myr for a top-heavy IMF
\citep{sch03}. The cause is directly related to the lifetimes of the
longest-lived (metal-free) star that is relevant for reionization in each
of the IMFs: a 10 $M_\odot$ star ($\sim$ 2 $\times 10^7$ yr) versus a 100
$M_\odot$ star ($\sim$ 3 $\times 10^6$ yr). The net difference in the
time-integrated ionizing radiation is a factor of only $\sim$ 1.4
\citep{bromm, sch02, sch03}.

Lastly, for the conversion from $\eta_{\rm Lyc}$ to $N_{\rm Lyc}/N_{\rm
b}$, we assume that $Z_{\rm IGM} \sim$ 10$^{-2.5} Z_\odot$, with the
correspondingly scaled levels of carbon and oxygen with respect to their
solar abundances\footnote{Note that the solar abundance values of some metals
have been significantly revised in recent years. Although this is currently
unresolved, we note that if we use, e.g., \citet{holweger}, the values of
$N_{\rm Lyc}/N_{\rm b}$ in this work will increase slightly for carbon and
decrease by about 30\% for oxygen.}  (\citealt{shull93}, and references
therein).  We take $\langle E_{\rm Lyc} \rangle$ to be 21 eV for $Z_\star =
Z_\odot$ and $Z_\star = 0.001$ stars, 27 eV for $Z_\star = 0$ stars in a
1--100 $M_\odot$ IMF, and 30 eV for $Z_\star = 0$ stars in a top-heavy IMF
\citep{sch03}.

\section{Results}

We first show that carbon and oxygen are mostly the products of IMSs and
massive stars respectively, and that IMSs are consequently significant
contributors to cosmological metal generation.  Figure 1 displays the total
IMF-weighted yields for metals, $^{12}$C and $^{16}$O as a function of
stellar mass for VMSs and for three values of $Z_\star$ for a present-day
IMF. Using the total ejected mass, $M_i$, in each element or metals as a
function of stellar mass $M$, the $y$-axis in the figure is calculated by
weighting each $M_i$ with the IMF as $M_i \phi(M)$.  Using the yields and
ionizing spectra as described in \S 2, we calculate $\eta_{\rm Lyc}$ and
$N_{\rm Lyc}/N_{\rm b}$ in Table 1 for several cases: (a) as a function of
$Z_\star$, (b) as defined with respect to total metal, $^{12}$C, and
$^{16}$O yield, and (c) with metal yields included from three mass ranges
(1--100 $M_\odot$, 8--100 $M_\odot$, and 100-1000 $M_\odot$).  In the
second case, the conversion from $\eta_{\rm Lyc}$ to $N_{\rm Lyc}/N_{\rm
b}$ involves two factors that will cause $N_{\rm Lyc}/N_{\rm b}$ to vary
within each column in the table: $\eta_{\rm Lyc}$ and $Z_{\rm IGM}$, which
will both vary at fixed $Z_\star$ depending on whether we are considering
$\eta_{\rm Lyc}$ and $N_{\rm Lyc}/N_{\rm b}$ with respect to total metals,
$^{12}$C, or $^{16}$O.  The last case is intended to evaluate the impact of
the metal contribution from various regions of the stellar IMF,
particularly from IMSs, on $\eta_{\rm Lyc}$ and $N_{\rm Lyc}/N_{\rm b}$.

The results in Table 1, although subject to the uncertainties in stellar
modelling, reveal several trends of interest to high-$z$ IGM studies.
These may be summarized as follows.

First, $\eta_{\rm Lyc}$ can vary significantly with $Z_\star$ and with the
element with respect to which it is defined. As anticipated for the reasons
outlined in \S 2, there is a strong gain in $\eta_{\rm Lyc}$ and $N_{\rm
Lyc}/N_{\rm b}$ as $Z_\star$ decreases, with $Z=0$ stars in a present-day
IMF being up to 10-20 times more efficient at generating ionizing radiation
per unit metal yield than the $Z=Z_\odot$ stars in \citet{madshull}.

Second, although the values of $\eta_{\rm Lyc}$ approximately match those of,
e.g., \cite{sch02}, when the yields from IMSs are excluded, we see that
IMSs are in fact a substantial source of metals, particularly carbon, and
should be included in such calculations. One possible exception might be at
very high redshifts, when IMSs may not have had the time yet to eject their
nucleosynthetic products. In this case, the original definition
\citep{madshull} of $\eta_{\rm Lyc}$ which does not include IMS metal
yields would be valid, and $\eta_{\rm Lyc}$ may be stated to increase with
decreasing $Z_\star$ and/or increasing redshift.

Third, $\eta_{\rm Lyc, O}$ is the most sensitive to $Z_\star$, increasing
strongly from solar-$Z_\star$ to metal-free stars, for a present-day IMF.
This directly results from the following three trends: (1) $^{16}$O and
ionizing radiation are produced dominantly by the IMF's massive stars, (2)
with decreasing $Z_\star$, the production of $^{16}$O decreases while that of
ionizing photons increases, and therefore, (3) their ratio $\eta_{\rm Lyc,
O}$, in comparison to $\eta_{\rm Lyc, Z}$ or $\eta_{\rm Lyc, C}$, shows the
strongest increase with declining $Z_\star$ for stars in a present-day IMF.

Fourth, the value of $\eta_{\rm Lyc}$ directly impacts the fraction of
baryons required to be converted into a first stars population in
order to influence reionization (e.g., \citealt{bromm, mirees97}).
One possible implication of the first result above is that a factor of
10--20 fewer baryons need be part of such a population for them to be
relevant for reionization alone (which requires only $N_{\rm
Lyc}/N_{\rm b} \sim$ 10), if such stars are not required to account
for $Z_{\rm IGM}$. In this case, reionization by Pop III stars in a
present-day IMF occurs late (since they should generate fewer ionizing
photons than indicated by Table 1), and such stars would therefore
make a negligible contribution to the carbon detected in the high-$z$
IGM, particularly at $z \ga 6$. An alternate interpretation of Table
1, assuming that the IGM is relatively uniformly enriched by volume,
is that $Z=0$ stars (1--100 $M_\odot$) created the IGM metals and
consequently, reionization may occur early.  This is because the
values of $N_{\rm Lyc}/N_{\rm b}$ ($\sim$ 30--220) generated by these
stars in association with the detected $Z_{\rm IGM}$ well exceeds that
required for a late H~I reionization at $z \sim 6$. If clusters of
such stars are responsible for both the reionization and metal
enrichment of the IGM, they should be easily detected by future
missions such as the JWST \citep{tsv,tgs}, and the amount of star
formation at $z \ga 9$ is potentially significant (\S 1).

Lastly, despite their high ionizing photon {\it rate}, VMSs are {\it less}
efficient at generating total ionizing radiation {\it per unit metal mass}
than are metal-free stars in a present-day IMF, owing to the greatly
boosted metal yield from a top-heavy IMF. (Both of these stellar
populations, however, have approximately equal $\eta_{\rm Lyc, C}$, with
the difference being largest when IMSs are excluded, or equivalently at $z
\ga 6$.)  Therefore, VMSs are not necessarily preferred as a more efficient
source of ionizing radiation at early epochs on nucleosynthetic grounds in
association with a given detected level of IGM enrichment, and possibly
based on reionization requirements as well (see, e.g.,
\citealt{somerville2}).  Furthermore, recent studies (e.g.,
\citealt{schneider}) indicate that the transition from a top-heavy to a
present-day IMF occurs at gas metallicities of about $10^{-4} Z_\odot$. If
this were true, then we can scale the numbers for VMSs in Table 1 down by
1.5 orders of magnitude, and derive that VMSs may produce only $\sim$ 0.35
ionizing photons per baryon for an IGM metallicity of $10^{-4} Z_\odot$
before they pollute their environments and cease forming. We emphasize,
however, that VMSs may still be important for reionization alone, given the
right combination of conditions, owing to their high ionizing rates.

The conversion between $\eta_{\rm Lyc}$ and $N_{\rm Lyc}/N_{\rm b}$ does
not account for the relative propagation timescales of metals and photons
from starforming regions to the IGM, which will impact the evolution of
IGM abundance ratios between individual elements, e.g., [C/O]. Furthermore,
the definition of $\eta_{\rm Lyc}$ implicitly assumes the instantaneous
creation of the metals. This is not necessarily true at high redshifts,
e.g., to generate sufficient carbon at $z \sim$ 3--6 requires the star
formation to have occurred at $z \ga$ 6--9.  As we noted earlier, this
becomes particularly problematic for the transport of the carbon to the IGM
through SN-driven winds.  We explore the cosmological relevance of these
timescale issues in more detail in a forthcoming paper.

\acknowledgements

We are grateful to Jason Tumlinson for valuable input and calculations.  We
thank Mike Shull and Jessica Rosenberg for many helpful discussions, and
Daniel Schaerer, Alessandro Chieffi, Oscar Straniero and Sara Ellison for
useful correspondence. We thank our anonymous referee for helpful
suggestions which improved the manuscript. A.~V. gratefully acknowledges
the support of NSF grant AST-0201670. J.~W.~T. acknowledges the support of
DOE grant DE-FG02-91ER40606 in Nuclear Physics and Astrophysics at The
University of Chicago.

\clearpage

\begin{figure}[!ht]
\plotone{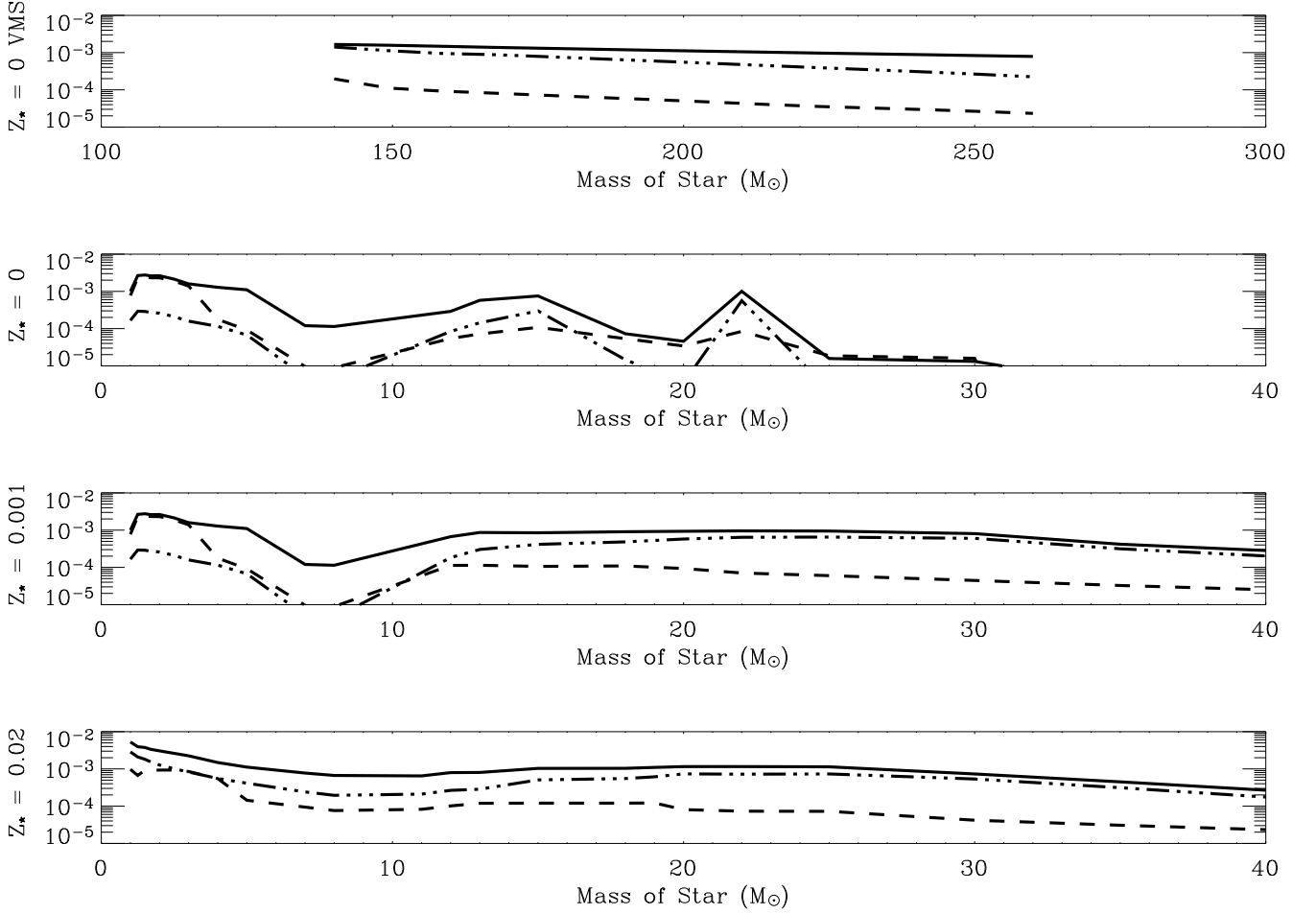}
\caption{IMF-weighted total ejected masses of metals (solid lines), carbon
  ($^{12}$C; dashed lines) and oxygen ($^{16}$O; dashed-dotted lines) as a
  function of stellar mass for a Salpeter IMF. Top panel: $Z_\star=0$ VMSs
  in the mass range 100--1000 $M_\odot$. Lower three panels: stars of
  metallicity 0, 0.001 and $Z_\odot = 0.02$ in the mass range 1--100
  $M_\odot$.}
\end{figure}

\begin{table*}[!hb]
\begin{center}
\caption{Ionizing Efficiency as a Function of Stellar Metallicity} 
\vspace{0.1in}
\begin{tabular}{lccccccc}
\tableline \tableline
  & $Z_\star$ = 0  & \multicolumn{2}{c}{$Z_\star$ = 0 (1--100
 $M_\odot$)} & \multicolumn{2}{c}{$Z_\star$ = 0.001
 (1--100 $M_\odot$)} & \multicolumn{2}{c}{$Z_\star$ = $Z_\odot$ (1--100
 $M_\odot$)}  \\ \tableline
 & ($10^2-10^3 M_\odot$) & $M_Z (\geq 1 M_\odot)$ &  $M_Z (\geq 8 M_\odot)$ &  $M_Z (\geq 1
 M_\odot)$ & $M_Z (\geq 8 M_\odot)$ & $M_Z (\geq 1 M_\odot)$ & $M_Z (\geq
 8 M_\odot)$ \\ \tableline
 &  & & & & & & \\
$\eta_{\rm Lyc, Z}$ ($N_{\rm Lyc}/N_{\rm b}$) & 0.005 (11) & 0.035 (81) & 0.08 (186) & 0.006 (18)  & 0.008 (24) & 0.003 (9) & 0.004 (13) \\
$\eta_{\rm Lyc, C}$ ($N_{\rm Lyc}/N_{\rm b}$)  & 0.098 (32)  & 0.086 (31) & 0.48 (173) & 0.025 (12) & 0.088 (41) & 0.02 (9) & 0.048 (22) \\
$\eta_{\rm Lyc, O}$ ($N_{\rm Lyc}/N_{\rm b}$) & 0.01 (10) & 0.15 (172)  & 0.2 (223) & 0.013 (18) & 0.013 (19) & 0.005 (8) & 0.007 (10) \\
\tableline
\end{tabular}
\vspace{-0.3in} 
\tablecomments{The conversion efficiency of rest mass in
metals, carbon and oxygen to ionizing radiation and the associated number
of ionizing photons per baryon, as a function of stellar metallicity,
$Z_\star$, and the mass range of the stellar population. Cases are also
shown with metal contributions from different regions of the stellar
IMF. All models assume a Salpeter slope for the IMF.}
\end{center}
\end{table*}

\end{document}